\documentclass[11pt,a4paper,english,showpacs,superscriptaddress]{revtex4-1}
\usepackage{amssymb}
\usepackage{amsmath}
\usepackage{graphicx}
\usepackage{babel}
\usepackage{hyperref}
\usepackage{color}

\begin{document}

\title{Dynamical generation of mass in the noncommutative supersymmetric Schwinger model}

\author{A.~C.~Lehum}
\email{andrelehum@ect.ufrn.br}
\affiliation{Escola de Ci\^encias e Tecnologia, Universidade Federal do Rio Grande do Norte\\
Caixa Postal 1524, 59072-970, Natal, Rio Grande do Norte, Brazil}

\author{J. R. Nascimento}
\email{jroberto@fisica.ufpb.br}
\affiliation{Departamento de F\'{\i}sica, Universidade Federal da Para\'{\i}ba\\
 Caixa Postal 5008, 58051-970, Jo\~ao Pessoa, Para\'{\i}ba, Brazil}

\author{A. Yu. Petrov}
\email{petrov@fisica.ufpb.br}
\affiliation{Departamento de F\'{\i}sica, Universidade Federal da Para\'{\i}ba\\
 Caixa Postal 5008, 58051-970, Jo\~ao Pessoa, Para\'{\i}ba, Brazil}


\begin{abstract}

Within the superfield formalism, we study the dynamical generation of mass to the gauge superfield in the noncommutative two-dimensional supersymmetric Schwinger model. We show that the radiatively generated mass for the gauge superfield does not depend on the noncommutative parameter $\Theta$ up to one-loop order.

\end{abstract}

\maketitle

\section{Introduction}

In the last years, field theory models constructed in lower dimensions of space-time have been intensively discussed, because, through AdS/CFT correspondence \cite{Maldacena1997}, they could be related to more elaborated theories in higher dimensions. The supersymmetric gauge theories in lower dimensions considered as candidates to describe $M2$-branes~\cite{Bagger:2006sk,Gustavsson:2007vu,Aharony:2008ug} attract main attention. Currently, a large number of papers is devoted to the study of several aspects of these theories, such as effective potential calculations~\cite{Gaiotto:2007qi,Benna:2008zy,Buchbinder:2010ez}, dualities~\cite{Karlhede:1986qd,Ferrari:2006vy,Gomes:2008pi,Ferrari:2008he} and generation of mass through the spontaneous symmetry breaking mechanism~\cite{Lehum:2007nf,Ferrari:2010ex,Lehum:2010tt,Gallegos:2011ux,Lehum:arxiv}.
The Schwinger model, i.e. quantum electrodynamics in two dimensions of space-time, is of the special interest among the low-dimensional gauge theories since it possesses the interesting feature of dynamical generation of mass and is known as an example of a confining model in the commutative~\cite{Abdalla:1991vua} and noncommutative space-time~\cite{Saha:2006qt}. It is worth it to mention that the two-dimensional noncommutative supersymmetric (SUSY) quantum electrodynamics is finite to all loop orders in perturbation theory \cite{Gomes:2011aa}, with the same conclusion being true for the three-dimensional commutative SUSY QED \cite{Ferrari:2007mh}.


Throughout this paper, we are using the superfield formalism; it is the more convenient way to evaluate Feynman diagrams in SUSY models. It preserves a manifest supersymmetry in all stages of calculations, avoiding potential problems, such as, for example, the lack of a supersymmetric renormalization presented in Ref.~\cite{Inami:2000eb} is not a problem when supergraph techniques are used~\cite{Ferrari:2006xx}. The present paper is organized as follows. In Sec. \ref{model}, we introduce the model and evaluate the quadratic part of the effective action for the gauge superfield, in order to observe the dynamical generation of mass to the noncommutative supersymmetric Schwinger model. We demonstrate that this effect is independent of the noncommutative parameter $\Theta$, up to one-loop order, just as the nonsupersymmetric version of present model~\cite{Ardalan:2010qb}. In Sec. \ref{final}, we present our last comments and remarks. 

\section{Noncommutative SUSY Schwinger model}\label{model}

\subsection{Pure gauge theory}

Our starting point is the classical action of the noncommutative SUSY Schwinger model,
\begin{eqnarray}\label{eq1}
S&=&\int{d^2xd^2\theta}\Big{\{}\frac{1}{2}W^{\alpha}*W_{\alpha}
-\frac{1}{4\xi} D^{\alpha}\Gamma_{\alpha}D^2D^{\beta}\Gamma_{\beta}+\dfrac{1}{2}\bar{c}D^{\alpha}\left(D_{\alpha}c-ie[\Gamma_\alpha,c]_*\right)\Big{\}}~,
\end{eqnarray}

\noindent
where $W^{\alpha}=\frac{1}{2}D^{\beta}D^{\alpha}\Gamma_{\beta}-\dfrac{ie}{2}[\Gamma^{\beta},D_{\beta}\Gamma_{\alpha}]_{*}-\dfrac{e^2}{6}[\Gamma^{\beta},\{ \Gamma_{\beta},\Gamma_{\alpha}\}_{*}]_{*}$ is the noncommutative gauge superfield strength which transforms covariantly, $W'_{\alpha}=e^{iK}*W_{\alpha}*e^{iK}$, with $K=K(x,\theta)$ being a real scalar superfield, and the exponential is treated in the sense of the Moyal star-product. Essentially, as discussed in Ref.~\cite{Gomes:2011aa}, there is no difference between conventions and notations for supersymmetric models defined in three and two dimensions of space-time. Therefore, we use the notations and conventions as adopted in Ref.~\cite{SGRS}. The inclusion of a gauge fixing and the corresponding Faddeev-Popov ghosts terms is required to quantize this model.

For later purposes, let us write the quadratic part of the gauge superfield action, which is given by
\begin{eqnarray}\label{eq1a}
S_2(gauge)&=&\int{d^2xd^2\theta}\Big{\{}-\frac{1}{8}\Gamma_{\gamma}D^{\alpha}D^{\gamma}D^{\beta}D_{\alpha}\Gamma_{\beta}
+\mathrm{gauge~fixing}\Big{\}}\nonumber\\
&=&\int{\frac{d^2p}{(2\pi)^2}}d^2\theta\Big{\{}-\frac{1}{4}\Gamma_{\gamma}(p,\theta)p^2\left(C_{\beta\gamma}+\frac{p_{\beta\gamma}D^2}{p^2}\right)\Gamma_{\beta}(-p,\theta)+\mathrm{gauge~fixing}\Big{\}}.
\end{eqnarray}

The propagators obtained from Eq.(\ref{eq1}), for the pure gauge sector, can be cast as
\begin{eqnarray}\label{eq4}
\langle \Gamma^{\alpha}(-p,\theta_1)\Gamma^{\beta}(p,\theta_2)\rangle&=&\frac{i}{2}\frac{D^2}{(p^2)^2}
\left(D_{\beta}D_{\alpha}-\xi D_{\alpha}D_{\beta}\right)\delta_{12}\nonumber\\
&=&\frac{i}{2}\frac{(1+\xi)C_{\beta\alpha}p^2+(1-\xi)p_{\beta\alpha}D^2}{(p^2)^2}
\delta_{12}~,\nonumber\\
\langle c(p,\theta_1)\bar{c}(-p,\theta_2)\rangle&=&i\frac{D^2}{p^2}\delta_{12}~,
\end{eqnarray}

\noindent
where $\delta_{12}=\delta^2(\theta_1-\theta_2)$. 

For simplicity, but without loss of generality, we will work in the Feynman gauge, i.e. we choose $\xi=1$.

\begin{figure}[htbp] \begin{center} \includegraphics[width={8cm}, angle=-90]{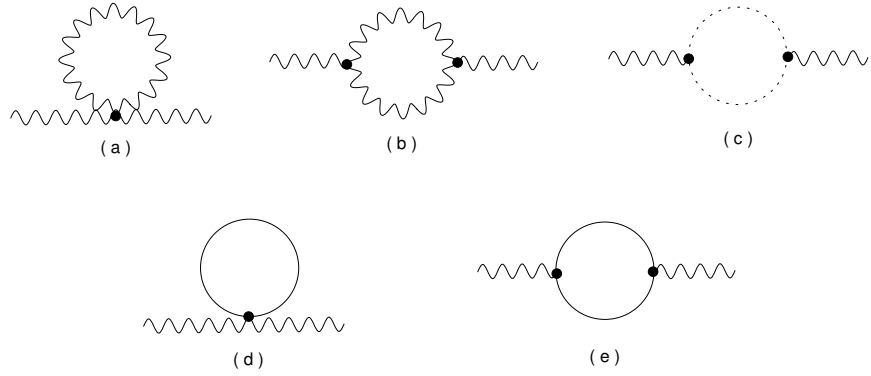}  \end{center} \caption{Diagrams which contribute with quadratic part of gauge superfield effective action. In this figure, continuous lines represent matter superfield propagators, wavy lines gauge superfield propagators, and dashed lines ghost superfield propagators.}\label{Fig:diagrams} \end{figure}

The effective action receives one-loop contributions from the diagrams drawn in Fig. \ref{Fig:diagrams}. Performing the D-algebra manipulations with the help of the Mathematica$^{\copyright}$ packet SusyMath \cite{Ferrari:2007sc}, we arrive at the following results. 
The supergraph \ref{Fig:diagrams}(a) is vanishing, while other contributions can be cast as
\begin{eqnarray}\label{eq5}
S_{\ref{Fig:diagrams}(b)}=&&-\frac{e^2}{4}\int{\frac{d^2p}{(2\pi)^2}}d^2\theta~\int{\frac{d^2k}{(2\pi)^2}}~\Gamma^{\alpha}(p,\theta)\frac{\left(p_{\alpha\beta}D^2+2C_{\beta\alpha}k^2\right)\sin^2{(k\wedge p)}}{k^2(k+p)^2}\Gamma^{\beta}(-p,\theta);
\end{eqnarray}
\begin{eqnarray}\label{eq6}
S_{\ref{Fig:diagrams}(c)}=&&-\frac{e^2}{4}\int{\frac{d^2p}{(2\pi)^2}}d^2\theta~\int{\frac{d^2k}{(2\pi)^2}}~\Gamma^{\alpha}(p,\theta)\frac{\left(p^2-2k^2\right)C_{\beta\alpha}\sin^2{(k\wedge p)}}{k^2(k+p)^2}\Gamma^{\beta}(-p,\theta).
\end{eqnarray}

Performing some algebraic manipulations and adding these two contributions, we have
\begin{eqnarray}\label{eq7}
S_{1loop}(gauge)=-\frac{e^2}{4}\int{\frac{d^2p}{(2\pi)^2}}d^2\theta~\Gamma^{\alpha}(p,\theta)\left(p_{\alpha\beta}D^2+C_{\beta\alpha}p^2\right)\Gamma^{\beta}(-p,\theta)~\int{\frac{d^2k}{(2\pi)^2}}\frac{\sin^2{(k\wedge p)}}{k^2(k+p)^2}.
\end{eqnarray}

Using Feynman representation and trivial transformations, we can rewrite the integral over $k$ as
\begin{eqnarray}\label{eq7a}
I=\frac{1}{2}\int\frac{d^2k}{(2\pi)^2}\int_0^1 dx\Big[\frac{1}{[k^2+p^2x(1-x)]^2}
-\frac{\cos(2k\wedge p)}{[k^2+p^2x(1-x)]^2}\Big]\equiv \frac{1}{2}(I_1-I_2),
\end{eqnarray}

\noindent where $I_1$ and $I_2$ are planar and nonplanar contributions.

It is easy to see that
\begin{eqnarray}\label{eq7b0}
I_1=\frac{1}{4\pi}\int_0^1 dx \frac{1}{M^2(x)},
\end{eqnarray}

\noindent where $M^2(x)=-p^2x(1-x)$. We note, however, that this integral diverges both at higher and lower limits, so, we implement the cutoff regularizations on both limits which yields
\begin{eqnarray}\label{eq7b}
I_{1\,reg}=\frac{1}{4\pi}\int_{\epsilon_1}^{1-\epsilon_2} dx \frac{1}{M^2(x)},
\end{eqnarray}
where, at the end of the calculations, one must put $\epsilon_1,\epsilon_2\to 0$.

Applying the results from Ref.~\cite{AlvarezGaume:2001ka} and regularizing the integral in the similar way, we find
\begin{eqnarray}\label{eq7c}
I_{2\,reg}
=\frac{1}{4\pi}\int_{\epsilon_1}^{1-\epsilon_2} dx \frac{1}{M^2(x)}\sqrt{4M^2(x)p\circ p}K_{-1}(\sqrt{4M^2(x)p\circ p}),
\end{eqnarray}

\noindent where $p\circ p\equiv p_m\Theta^{mn}\Theta_{nl}p^l$, which in two dimensions of space-time can be written as $\Theta^2p^2$, once we assume $\Theta^{mn}=\Theta\epsilon^{mn}$. Thus, we arrive at the final expression for the integral:
\begin{eqnarray}\label{eq7d}
I_{reg}=\frac{1}{8\pi}\int_{\epsilon_1}^{1-\epsilon_2} dx \frac{1}{M^2(x)}\left[1-\sqrt{4M^2(x)\Theta^2p^2}K_{-1}(\sqrt{4M^2(x)\Theta^2p^2})
\right].
\end{eqnarray}

Unfortunately, this integral cannot be more simplified since the modified Bessel function cannot be expressed in terms of the elementary functions. However, for our aims, that is, for studying the mass generation, we can use its asymptotic behavior for small and large arguments: 
\begin{eqnarray}\label{bessel}
K_{\pm 1}(s\rightarrow0 ) &=&\frac{1}{s}+cs+\mathcal{O}(s^2),\nonumber\\
K_{\nu}(s\rightarrow\infty ) &=&\sqrt{\frac{2}{\pi s}}~{\mathrm{e}}^{-s},\nonumber
\end{eqnarray}

\noindent
where $c$ is a constant whose explicit value is not important. 

Therefore, in the limit of $\Theta^2\rightarrow 0$, we obtain that the integral $I$ is of order of $\Theta^2 p^2$. Taking into account only two leading terms of the expansion of the modified Bessel function, we find that the terms singular in the limits $\epsilon_1\to 0$ and $\epsilon_2\to 0$ turn out to be completely cancelled, after which we can remove the regularization, and the integral over the Feynman parameter $x$ is trivial. We note that had we used other regularization, for example, the dimensional one, the situation could be just the same, that is, the final result would be free of any singularities. As a result, we find that the effective action is just given by
\begin{eqnarray}\label{eq7e}
S_{eff}(gauge)=&&-\frac{1}{4}\int{\frac{d^2p}{(2\pi)^2}}d^2\theta~\Gamma^{\alpha}(p,\theta)p^2\left(C_{\beta\alpha}+\frac{p_{\alpha\beta}D^2}{p^2}\right)\left[1+ce^2\Theta^2p^2\right]  \Gamma^{\beta}(-p,\theta).
\end{eqnarray}

We conclude that no mass is dynamically generated due to a self-interacting gauge sector in the small noncommutativity limit. 

It is clear that this result differs from that obtained in the large noncommutativity limit~\cite{Ambjorn:1999ts,Aoki:1999vr,Ambjorn:2000nb,Szabo:2001kg,Bietenholz:2004as} where the noncommutative $U(1)$ model was shown to be equivalent to the commutative $U(N)$ theory in the large-$N$ limit. The reason is as follows: while the noncommutative QED in the limit $\Theta\rightarrow\infty$ behaves like a $U(N)$ Yang-Mills theory in the large-$N$ limit, in the limit $\Theta\rightarrow0$, the noncommutative QED behaves like a free theory. Therefore the dynamics of this theory in these two limits radically differs. Hence, the results for one limit should not be expected to be reproduced in another limit.

On the other hand, considering the theory in the limit of $\Theta^2\rightarrow \infty$, we obtain that the integral $I$ behaves like $I_1$. To study dynamical generation of mass, it is enough to evaluate the effective action in the limit $p^2\rightarrow 0$~\cite{Ardalan:2010qb}.  Performing the integral $\int{\frac{d^2k}{(2\pi)^2}}\frac{1}{k^2(k+p)^2}$ with the help of an infrared regulator and taking $p^2\rightarrow 0$, we obtain the following effective action
\begin{eqnarray}\label{eq7f}
S_{eff}(gauge)&=&-\frac{1}{4}\int{\frac{d^2p}{(2\pi)^2}}d^2\theta~\Gamma^{\alpha}(p,\theta) p^2 \left(C_{\beta\alpha}+\frac{p_{\alpha\beta}D^2}{p^2}\right) \left[1+\frac{e^2}{3\sqrt{3}p^2}\right]  \Gamma^{\beta}(-p,\theta)\nonumber\\
&=&-\frac{1}{4}\int{\frac{d^2p}{(2\pi)^2}}d^2\theta~\Gamma^{\alpha}(p,\theta)  \left(C_{\beta\alpha}+\frac{p_{\alpha\beta}D^2}{p^2}\right) \left[p^2+M_\gamma^2\right]  \Gamma^{\beta}(-p,\theta),
\end{eqnarray}

\noindent 
where we find the presence of a massive pole for the perturbative full propagator arisen from the effective action (\ref{eq7f}), with $M_{\gamma}^2=\dfrac{e^2}{3\sqrt{3~}}$, corroborating the results obtained in nonsupersymmetric models~\cite{Ambjorn:1999ts,Aoki:1999vr,Ambjorn:2000nb,Szabo:2001kg,Bietenholz:2004as}. One should notice that the dynamically generated mass is independent of noncommutative parameter $\Theta$.




\subsection{Matter superfields in fundamental representation}

The form of the matter couplings depends on assumed noncommutative representation for matter superfields. Let us first consider matter superfields in the fundamental left-representation. To the action (\ref{eq1}), we add the following matter superfield action:
\begin{eqnarray}\label{eq1aa}
S\!=\!\!\int\!{d^2xd^2\theta}\Big{\{}\!\!-\bar\Phi D^2\Phi-\frac{e^2}{2}\bar\Phi*\Gamma^{\alpha}*\Gamma_{\alpha}*\Phi
+i\frac{e}{2}\big[D^{\alpha}\bar{\Phi}*\Gamma_{\alpha}*\Phi-\bar\Phi* \Gamma^{\alpha}*D_{\alpha}\Phi~\big]\Big{\}},
\end{eqnarray}

\noindent from which we obtain the matter superfield propagator given by
\begin{eqnarray}\label{eq4a}
\langle \Phi(k,\theta_1)\bar\Phi(-k,\theta_2)\rangle&=&-i\frac{D^2}{k^2}\delta_{12}~.
\end{eqnarray}
 
The contributions due to matter coupling, Figs. \ref{Fig:diagrams} (d) and (e), in the fundamental representation, can be cast as
\begin{eqnarray}\label{eq8}
S_{\ref{Fig:diagrams}(d+e)}&=&-\frac{e^2}{2}\int{\frac{d^2p}{(2\pi)^2}}d^2\theta\int{\frac{d^2k}{(2\pi)^2}}\Gamma^{\alpha}(p,\theta)\Big{\{}\frac{C_{\beta\alpha}}{k^2}-\frac{C_{\beta\alpha}}{(k+p)^2}+\nonumber\\
&+&\frac{1}{4}
\frac{\left(p^2C_{\beta\alpha}-p_{\beta\alpha}D^2\right)}{(k+p)^2k^2}\Big{\}}\Gamma^{\beta}(-p,\theta).
\end{eqnarray}

We can note that the logarithmic divergent terms cancel between each other, and the contribution to the quadratic part of effective action for the gauge superfield coming from the matter sector is given by
\begin{eqnarray}\label{eq8a}
S_{\ref{Fig:diagrams}(d+e)}&=&-\frac{e^2}{8}\int{\frac{d^2p}{(2\pi)^2}}d^2\theta~\Gamma^{\alpha}(p,\theta)\Big{\{}\int{\frac{d^2k}{(2\pi)^2}}\frac{\left(p^2~C_{\beta\alpha}+p_{\beta\alpha}D^2\right)}{(k+p)^2k^2}\Big{\}}\Gamma^{\beta}(-p,\theta).
\end{eqnarray}

Summing up the classical and quantum parts of the effective action in the small noncommutativity limit, we have
\begin{eqnarray}\label{eq8b}
S_{eff}&=&-\frac{1}{4}\int{\frac{d^2p}{(2\pi)^2}}d^2\theta~\Gamma^{\alpha}(p,\theta)~p^2\Big(C_{\beta\alpha}+\frac{p_{\beta\alpha}D^2}{p^2}\Big)\Big[1+
\frac{e^2}{2}\int{\frac{d^2k}{(2\pi)^2}}\frac{1}{k^2(k+p)^2}\Big]\Gamma^{\beta}(-p,\theta).
\end{eqnarray}



Performing the integral $\int{\frac{d^2k}{(2\pi)^2}}\frac{1}{k^2(k+p)^2}$ as was done before, we obtain 
\begin{eqnarray}\label{eq8c}
S_{eff}&=&-\frac{1}{4}\int{\frac{d^2p}{(2\pi)^2}}d^2\theta~\Big{\{}\Gamma^{\alpha}(p,\theta)~\left(C_{\beta\alpha}+\frac{p_{\beta\alpha}D^2}{p^2}\right)\left[p^2+M_{\gamma}^2\right]\Gamma^{\beta}(-p,\theta)\Big{\}},
\end{eqnarray}

\noindent
where we find again the presence of a massive pole for the perturbative full propagator, with $M_{\gamma}^2=\dfrac{e^2}{3\sqrt{3~}}$.  

Once the diagrams related to the matter contributions are planar, for matter in fundamental representation, the correction to the generated mass is valid for both cases, large and small noncommutativity. Therefore, the effective action in the limit $\Theta\rightarrow\infty$ can be cast as 
\begin{eqnarray}\label{eq8d}
S_{eff}-\frac{1}{4}\int{\frac{d^2p}{(2\pi)^2}}d^2\theta~\Big{\{}\Gamma^{\alpha}(p,\theta)~\left(C_{\beta\alpha}+\frac{p_{\beta\alpha}D^2}{p^2}\right)\left[p^2+M_{\gamma} ^{\prime 2} \right]\Gamma^{\beta}(-p,\theta)\Big{\}},
\end{eqnarray}

\noindent
where $M_{\gamma}^{\prime 2}=2 M_{\gamma}^2$.

\subsection{Matter superfields in adjoint representation}

When matter superfields are assumed to be in the noncommutative adjoint representation, the matter superfield action turns out to be
\begin{eqnarray}\label{eq1b}
S\!=\!\!\int\!\!{d^2xd^2\theta}\Big{\{}\!\!-\bar\Phi D^2\Phi+\frac{e^2}{2}[\bar\Phi,\Gamma^{\alpha}]_* *[\Gamma_{\alpha},\Phi]_*
-\frac{ie}{2}\left(D^{\alpha}\bar{\Phi}*[\Gamma_{\alpha},\Phi]_*-[\bar\Phi,\Gamma^{\alpha}]_* *D_{\alpha}\Phi\right)\Big{\}}~.
\end{eqnarray}

The vertices of interaction written in terms of noncommutative Moyal phases are given by Eqs. (\ref{eq:v6}) and (\ref{eq:v7}). For the adjoint representation, all couplings vanish in the commutative limit and the theory turns to be free. This sector contributes to effective action with~\cite{Gomes:2011aa}
\begin{eqnarray}\label{eq11}
S_{\ref{Fig:diagrams}(d+e)}=&&-\frac{e^2}{2}\int{\frac{d^2p}{(2\pi)^2}}d^2\theta~\Gamma^{\alpha}(p,\theta)\left(p_{\alpha\beta}D^2+C_{\beta\alpha}p^2\right)\Gamma^{\beta}(-p,\theta)\int{\frac{d^2k}{(2\pi)^2}}\frac{\sin^2{(k\wedge p)}}{k^2(k+p)^2}.
\end{eqnarray}

Adding Eq.(\ref{eq11}) with the contribution which comes from the gauge sector (\ref{eq7}), the quantum correction to the quadratic part of the gauge superfield effective action is given by
\begin{eqnarray}\label{eq12}
S_{1loop}&=&-\frac{3}{4}e^2~\int{\frac{d^2p}{(2\pi)^2}}d^2\theta~\Gamma^{\alpha}(p,\theta)p^2\left(C_{\beta\alpha}+\frac{p_{\beta\alpha}D^2}{p^2}\right)\int{\frac{d^2k}{(2\pi)^2}}\frac{\sin^2{(k\wedge p)}}{(k+p)^2k^2}\Gamma^{\beta}(-p,\theta)~.
\end{eqnarray}

In the limit of $\Theta\rightarrow 0$, one-loop quantum effects (\ref{eq12}) do not change the dynamics of the model when matter superfields are in the adjoint representation. Just as the pure gauge sector, cf. Eq.(\ref{eq7e}), the one-loop contribution is of the order $\Theta^2p^2$, and no generation of mass is present in this version of the model up to this order. This is related with the effect that in the commutative limit, the theory in the adjoint representation behaves like a free one.

In the opposite case, i.e. $\Theta\rightarrow\infty$, the effective action can be cast as
\begin{eqnarray}\label{eq12a}
S_{eff}&=&-\frac{1}{4}\int{\frac{d^2p}{(2\pi)^2}}d^2\theta~\Big{\{}\Gamma^{\alpha}(p,\theta)~\left(C_{\beta\alpha}+\frac{p_{\beta\alpha}D^2}{p^2}\right)\left[p^2+M_{\gamma}^{\prime\prime 2}\right]\Gamma^{\beta}(-p,\theta)\Big{\}},
\end{eqnarray}

\noindent
where $ M_{\gamma}^{\prime\prime 2} = 3M_{\gamma}^{2} $.

\section{Final remarks}\label{final}

In this paper, we have computed the effective action of the noncommutative gauge superfield in interaction with matter scalar superfield, both in fundamental and adjoint representations, in the noncommutative supersymmetric quantum electrodynamics in two-dimensional space-time, i.e., the SUSY Schwinger model. 

In the limit of small noncommutativity, i.e. $\Theta\sim0$, we observe that the model, with the matter in noncommutative fundamental representation, presents a dynamical generation of mass to the gauge superfield, which is an effect independent of the noncommutative parameter $\Theta$, up to one-loop order. When the matter superfields are in the noncommutative adjoint representation, the model does not exhibit such an effect in the small $\Theta$ limit. 

On the other hand, in the limit of large noncommutativity, i.e. $\Theta\rightarrow\infty$, we observe that the model presents a dynamical generation of mass to the gauge superfield with or without matter couplings. In respect of the pure gauge sector, the dynamical generation of mass is an effect related to the equivalence between the noncommutative $U(1)$ model and the commutative $U(N)$ Yang-Mills theory in the large-$N$ limit~\cite{Ambjorn:1999ts,Aoki:1999vr,Ambjorn:2000nb,Szabo:2001kg,Bietenholz:2004as}. 

As recently suggested in Ref.~\cite{Armoni:2011pa}, we expect that a $\Theta$ dependence in the generated mass can occur when three-loop Feynman supergraphs are taken into account. Actually, this work is in progress. Also, we expect that this approach can be useful for the study of the non-Abelian extension of the Schwinger model and for studies of the three-dimensional noncommutative SUSY QED. We are going to discuss these problems in forthcoming papers. Finally, it is well-known that space-time noncommutativity can break unitarity~\cite{unit}. In principle, this problem can be solved with use of the approach proposed in the papers~\cite{topt,topt1}. We expect that applying this formalism to this model should not give essentially different results.


\vspace{1cm}

{\bf Acknowledgements.} 
This work was partially supported by Conselho Nacional de Desenvolvimento Cient\'{\i}fico e Tecnol\'{o}gico (CNPq) and Funda\c{c}\~{a}o de Apoio \`{a} Pesquisa do Estado do Rio Grande do Norte (FAPERN). The work by A. C. Lehum and A. Yu. P. has been supported by the CNPq project No. 303392/2010-0 and 303461/2009-8, respectively.

\appendix

\section{Noncommutative vertices}\label{ncvertices}

The interaction vertices of noncommutative extensions of field theories are characterized by the presence of a noncommutative phase; that is, a function dependent of noncommutative parameter $\Theta$. These noncommutative vertices, for the model under consideration, are drawn in Fig. \ref{Fig:vertices} and their respective expressions are presented in the subsections below.

\subsection{Gauge superfield self-interactions and ghost couplings}\label{ncvertices1}

The noncommutative vertices for the gauge superfield self-interaction and Fadeev-Popov ghost couplings are given by:
\begin{eqnarray}\label{eq:v1}
V_{\ref{Fig:vertices}(a)}=\dfrac{e}{2}\sin{(k_2\wedge k_3)}~D^{\gamma}D^{\alpha}\Gamma_{\gamma}(k_1)\Gamma^{\beta}(k_2)D_{\beta}\Gamma_{\alpha}(k_3)~;
\end{eqnarray}
\begin{eqnarray}\label{eq:v2}
V_{\ref{Fig:vertices}(b)}=&&\dfrac{e^2}{2}\sin{(k_3\wedge k_4)}\sin{[k_2\wedge(k_3+k_4)]}\Big{\{}~\Gamma^{\gamma}(k_1)D_{\gamma}\Gamma^{\alpha}(k_2)\Gamma^{\beta}(k_3)D_{\beta}\Gamma_{\alpha}(k_4)\nonumber\\
&&+\dfrac{2}{3}~D^{\gamma}D^{\alpha}\Gamma_{\gamma}(k_1)\Gamma^{\beta}(k_2)\Gamma_{\beta}(k_3)\Gamma_{\alpha}(k_4)\Big{\}}~;
\end{eqnarray}
\begin{eqnarray}\label{eq:v3}
V_{\ref{Fig:vertices}(c)}=e\sin{(k_3\wedge k_2)}~\bar{c}(k_1)D^{\alpha}\left[\Gamma_{\alpha}(k_2)c(k_3)\right]~,
\end{eqnarray}

\noindent
where $a\wedge b=a_\mu b_\nu \Theta^{\mu\nu}$.

\subsection{Matter superfield couplings: fundamental representation}\label{ncvertices2}

When matter superfield is in the fundamental left representation, the noncommutative vertices can be cast as 
\begin{eqnarray}\label{eq:v4}
V_{\ref{Fig:vertices}(d)}=\dfrac{ie}{2}\left[e^{ik_3\wedge k_2}D^{\alpha}\bar\Phi(k_1)\Gamma_{\alpha}(k_2)\Phi(k_3)
-e^{ik_2\wedge k_3}\bar\Phi(k_1)\Gamma_{\alpha}(k_2)D^{\alpha}\Phi(k_3)\right]~;
\end{eqnarray}
\begin{eqnarray}\label{eq:v5}
V_{\ref{Fig:vertices}(e)}=-\dfrac{e^2}{2}e^{-i[k_2\wedge(k_3+k_4)+k_3\wedge k_4]}\Gamma^{\alpha}(k_1)\Gamma_{\alpha}(k_3)\bar{\Phi}(k_2) \Phi(k_4)~.
\end{eqnarray}

\subsection{Matter superfield couplings: adjoint representation}\label{ncvertices3}

\begin{figure}[htbp] \begin{center} \includegraphics[width={8cm},angle=-90]{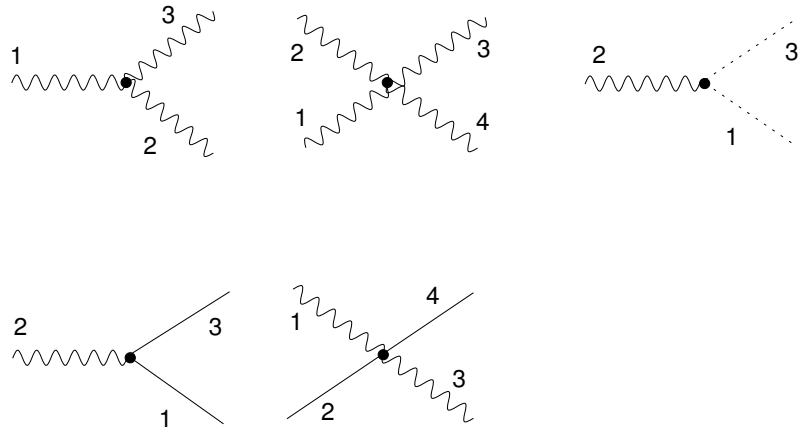}  \end{center} \caption{Noncommutative vertices. In this figure, continuous lines represent external matter, wavy lines external gauge, and dashed lines external ghost superfields.}\label{Fig:vertices} \end{figure}

When matter superfield is in the adjoint representation, the noncommutative vertices look like
\begin{eqnarray}\label{eq:v6}
V_{\ref{Fig:vertices}(d)}=\dfrac{e}{2}\sin{(k_2\wedge k_3)}\Gamma^{\alpha}(k_2)D_{\alpha}\left[\bar\Phi(k_1)\Phi(k_3)\right]~;
\end{eqnarray}
\begin{eqnarray}\label{eq:v7}
V_{\ref{Fig:vertices}(e)}=-2e^2\sin{[k_2\wedge(k_3+k_4)]}\sin{(k_3\wedge k_4)}\bar\Phi(k_1)\Gamma^{\alpha}(k_2)\Gamma_{\alpha}(k_3)\Phi(k_4)~.
\end{eqnarray}



\end{document}